\documentclass[sigconf,nonacm]{acmart}

\usepackage{algorithm}
\usepackage{algorithmic}
\usepackage{appendix}
\usepackage{amsmath}
\usepackage{amsfonts}
\usepackage{array}
\usepackage{booktabs}
\usepackage{caption}
\usepackage{comment}
\usepackage{csquotes}
\usepackage[us,12hr]{datetime}
\usepackage{enumitem}
\usepackage{etoolbox}
\usepackage{float}
\usepackage{geometry}
\usepackage{graphicx}
\usepackage{hyperref}
\usepackage{layouts}
\usepackage{multirow}
\usepackage{nameref}
\usepackage[percent]{overpic}
\usepackage{pdflscape}
\usepackage{rotating}
\usepackage{subfig}
\usepackage{tabularx}
\usepackage{textcomp}
\usepackage{wasysym}
\usepackage{xparse}

\makeatletter
\renewenvironment*{displayquote}
{\begingroup\setlength{\leftmargini}{0.3cm}\csq@getcargs{\csq@bdquote{}{}}}
{\csq@edquote\endgroup}
\makeatother

\apptocmd{\sloppy}{\hbadness 10000\relax}{}{}

\hypersetup{%
    colorlinks=true, 
    breaklinks=true, 
    urlcolor=blue, 
    linkcolor=blue, 
    citecolor=blue, 
}

\begin{document}
\pagestyle{plain}

\title{Systematic Analysis and Comparison\\of Security Advice as Datasets}

\author{Christopher Bellman}
\affiliation{  
    \city{Carleton University, Ottawa}
    \country{Canada}}
\email{chris@ccsl.carleton.ca}

\author{Paul C. van Oorschot}
\affiliation{
    \city{Carleton University, Ottawa}
    \country{Canada}}
\email{paulv@scs.carleton.ca}

\renewcommand{\shortauthors}{Bellman and Van Oorschot.}

\begin{abstract}
    A long list of documents have been offered as security advice, codes of practice, and security guidelines for building and using security products, including Internet of Things (IoT) devices. To date, little or no systematic analysis has been carried out on the advice datasets themselves. Towards addressing this, with IoT as a case study, we begin with an informal analysis of two documents offering advice related to IoT security---the ETSI Provisions and the UK DCMS Guidelines---and then carry out what we believe is the first systematic analysis of these advice datasets. Our analysis explains in what ways the ETSI Provisions are a positive evolution of the UK DCMS Guidelines. We also suggest aspects of security advice warranting special attention by those offering security advice. Such parties may find the systematic analysis method, which categorizes advice into predefined categories, to be of general interest beyond IoT itself.

    {\let\thefootnote\relax\footnote{{Journal version appears in Computers \& Security (Jan 2023). DOI: \url{https://doi.org/10.1016/j.cose.2022.102989}}. \copyright 2022. This manuscript version is made available under the CC-BY-NC-ND 4.0 license \url{https://creativecommons.org/licenses/by-nc-nd/4.0/}.}}
\end{abstract}


\maketitle

\section{Introduction}
\label{sec:introduction}
Over the past two decades, the Internet of Things (IoT) has progressed from ideas to products to ubiquitous deployment. Among retail consumers, IoT devices for so-called smart homes have become popular.  These consumer devices are typically controlled over networks, both locally and remotely, e.g., via smartphones.  Example devices include smart lightbulbs, smart door locks, smart thermostats, web cameras, and smart speakers.  

Perhaps unsurprisingly, given their network reachability, many of these devices have suffered from rather serious security vulnerabilities \cite{Alrawi2019, Kolias2017}.
With broad awareness of this state of affairs, over the past five to eight years in particular, a substantial number of organizations, including governments and industry associations, have offered IoT security guidelines and related advice documents, aiming to improve the state of affairs.  

However, the advice documents themselves appear to have received little analysis, and many
questions can be asked about the advice items they offer.  (We view the advice items as datasets in the context of our analysis.) Is the advice understandable by target audiences, e.g., IoT device manufacturers? Are there other target audiences, and what expertise is expected of them? Is advice intended to be \textit{actionable}, in the sense of specifying step-by-step details to follow, enabling reliable execution by advice recipients; or does it aim only to state desired end-goals or outcomes?

These questions are related to usability, but typically for IT workers other than end-users; the primary advice targets appear to be IoT manufacturers and software developers---but we find, typically not security experts, and expect few small to medium IoT device manufacturers employ deep teams of security experts. Note that here, the subject of analysis is security advice itself, not the resulting products per se. We believe that the quality and usability of such security advice is under-studied---both for IoT, and in general.   

In this work we provide an informal and then a systematic analysis of advice from two documents addressing IoT security \cite{DCMScodeOfPractice, ETSI2020}. For the systematic analysis we employ our \textit{security advice coding} (SAcoding) method, which involves qualitative data coding and has been used on a 1013-item advice dataset in previous work \cite{Barrera2022}. The datasets we explore herein are coarser (higher level), and our analysis includes comparing them. The SAcoding method measures actionability (Table~\ref{tab:definitions}), which we view as a desirable characteristic of advice. We also consider how security advice in general might be improved. 

Through this work and by analysis of these particular datasets, we aim to further understand various characteristics of popular security advice---hoping that this may illustrate characteristics important to advice-givers aiming to produce more well-targeted, actionable advice for advice recipients to rely upon. 

\section{Document Summaries}
\label{sec:background/summaries}
We first describe the relevant documents. Our aim is to provide an analysis and comparison of security advice datasets from two primary documents; we specifically focus on their primary advice content (rather than preamble and non-advice appendices). In what follows, DCMS refers to the UK government's Department for Digital, Culture, Media and Sport; ETSI is short for the European Telecommunications Standards Institute. 

The first primary document, which we call the \textit{DCMS 13 Guidelines} (from October 2018), is officially the DCMS \textit{Code of Practice for Consumer IoT Security} \cite{DCMScodeOfPractice}. It consists of 13 IoT security guidelines.  

The second primary document, which we refer to as the \textit{ETSI Provisions} (from June 2020), is \label{page:ETSIprovisionsdoc} ETSI's \textit{Cyber Security for Consumer IoT: Baseline Requirements}  \cite{ETSI2020}. It  uses each of the high-level DCMS 13 guidelines as category headings, with minor editorial changes. Under these headings, it provides a larger, finer-grained set of items called ``provisions'' for advice recipients to follow.

\label{sec:background/DCMSmapping}
Related to both of these primary documents is a large collection of advice  \cite{IoTSecMap} which we call the \textit{DCMS 1013-item dataset}, and view as representative of current IoT security advice. This is a compilation of items of IoT security advice, pruned down (by removing identical items) from a slightly larger original set. The advice items in this dataset are referenced in a third document, the \textit{DCMS mapping document} \cite{DCMS1}, which categorizes each of the 1013 items into one of the DCMS 13 guidelines that best suits it. 

The mapping document is positioned as a ``reference and tool for users of the Code of Practice [DCMS 13 Guidelines]'' \cite{DCMS1}, suggesting that those looking to follow each guideline would consult this mapping document for how to find technical details supporting each. Likewise, the ETSI Provisions document suggests use of the DCMS 1013-item dataset for technical details, albeit without explaining how readers are expected to use it. 

We now give more background on the two primary documents.

\subsection{DCMS 13 Guidelines}
\label{sec:background/DCMSguidelines}
    
The DCMS 13 Guidelines define 13 guidelines designed for pre-deployment stakeholders to use for improving the security of their products and services. These guidelines (examples are in Appendix, Table \ref{tab:taggingoutput/DCMSSubTopic}) are positioned as ``practical steps'', and also as ``outcome-focused'', giving stakeholders the flexibility to follow each guideline on their own terms rather than offering specific means to execute them \cite{DCMScodeOfPractice}. (This leads to a contradiction, as we discuss shortly.)
    
As such, these are high-level guidelines (i.e., lacking in technical detail) that cover a wide variety of IoT security topics. The first three guidelines are intentionally ordered first to signal their priority as those that the DCMS suggests may have the greatest and immediate security impact for a stakeholder. Each of the 13 guidelines follow a basic four-part structure as we now summarize.     

\paragraph{(1) Guideline Title.}
\label{sec:background/DCMSguidelines/structure/title}
A short title conveys each guideline's primary topic and general direction, compacted to fit on one line. The title is not always representative of the full recommendation given in the longer description. We refer to the  title as the \textit{guideline category}, generally representing the guideline's theme.
    
\paragraph{(2) Guideline Description.}
\label{sec:background/DCMSguidelines/structure/pinkbox}
For each guideline, a pink box highlights the core guideline text. The description conveys the key part of each guideline (or entire guideline if short), expressing what is intended for the advice follower to achieve. This text is the primary focus of our analysis. For example, DCMS-5 has description: 

\begin{displayquote}
    \textit{Security-sensitive data, including any remote management and control, should be encrypted in transit, appropriate to the properties of the technology and usage. All keys should be managed securely.}
\end{displayquote}

\noindent The description apparently aims to express the goal of the guideline more than \textit{how} to follow it, consistent with the document positioning the guidelines as
``\textit{outcome-focused, rather than prescriptive}'' \cite{DCMScodeOfPractice}, which suggests to us non-actionability.\footnote{
    Table~\ref{tab:definitions} gives explicit definitions of a few basic terms as used herein.
} However, this appears to contradict other positioning of the guidelines as ``\textit{practical steps}'', which we associate with actionability.
    
\paragraph{(3) Further Description/Explanation.}
\label{sec:background/DCMSguidelines/structure/explanation}
Supporting text accompanying each guideline gives a non-technical explanation of rationale, why the guideline is important, or what common problem it addresses. In a few cases this text notes additional avenues to solving a general problem. In some cases it expands on the guideline description; in others it goes off in independent directions.
    
\paragraph{(4) Guideline Footnote.}
\label{sec:background/DCMSguidelines/structure/footnote}
Two of the 13 guidelines have a brief footnotes section giving a short list of references indicating external sources for supplementary information, or further clarification of some aspect (e.g., the meaning of ``competent industry bodies'' \cite{DCMScodeOfPractice}, mentioning GSMA \cite{GSMA} and the IoT Security Foundation \cite{IoTSF}).
    
The document's end has more explanation for 6 of the guidelines, aiming to address frequent questions and positioning these as ``additional explanatory notes'' \cite{DCMScodeOfPractice}, perhaps suggesting that they are not needed in order to carry out the main advice. These sometimes include further advice or context on why the guideline is important. 
    
\subsection{ETSI Provisions}
\label{sec:background/ETSI}
The ETSI Provisions document  extends the 13  categories from the DCMS 13 Guidelines, and appears to be recognized by the DCMS as an extension of their work (DCMS lists it in their \textit{Secure by Design} project \cite{SecureByDesign}, and contributes a subset of the content, notably the introductory material and guideline categories).

While portions of these two documents are similar, by our reading, the DCMS 13 Guidelines and ETSI Provisions differ in scope. The DCMS guidelines appear more as high-level advice to be followed at the discretion of a manufacturer \cite{DCMScodeOfPractice}, while the ETSI provisions are positioned \cite{ETSI2020} as ``technical controls''  with the intent to be measurable for compliance when combined with technical details such as those in the DCMS 1013 dataset, or advice from ENISA \cite{ENISA1}, the IoT Security Foundation \cite{IOTSF2}, and the GSMA \cite{GSMAGuidelines}.

The ETSI Provisions document is built around the DCMS' 13 guideline categories, plus an extra guideline category (on how to protect user data processed by an IoT device).  We include this in our analysis, but exclude a further new section on how to report on implementation of provided advice (we view that as separate from advice items/categories per se).  

Each category in the ETSI Provisions document has an optional category description (a few sentences on the category goal or scope) and from 1 to 16 ``provisions''. Each provision has: (1) a one-sentence description of what is required by the advice follower or implementer. It may also have: (2) further description of the provision; and (3) examples or notes of where or how the provision would be applied, as context. Many provisions include all three; some have only the first. 

\section{Informal Comparison and Critique}
Before a systematic comparison, we first informally compare and critique the DCMS 13 Guidelines and ETSI Provisions documents---expecting the latter to be an improvement, as an evolution. As we do so, we begin to characterize what we recognize as aspects of advice that make it more (or less) actionable. 

\subsection{Positioning of Documents}
\label{sec:criticism/positioning}
We first distill our view of the stated purpose and positioning of each document. A first criticism is that the positioning of both documents is unclear, and by our reading, contradictory in places. 

The DCMS 13 Guidelines give four statements of positioning \cite{DCMScodeOfPractice}: 
        
\begin{displayquote}
    \textit{(D1) This Code of Practice sets out practical steps for IoT manufacturers and other industry stakeholders to improve the security of consumer IoT products and associated services.}
\end{displayquote}
        
\begin{displayquote}
    \label{page:d2}
    \textit{(D2) The guidelines bring together what is widely considered good practice in IoT security. They are outcome-focused, rather than prescriptive, giving organisations the flexibility to innovate and implement security solutions appropriate for their products.}
\end{displayquote}
        
\begin{displayquote}
    \textit{(D3) A number of industry bodies and international fora are developing security recommendations and standards for IoT. This Code of Practice is designed to be complementary to and supportive of those efforts and relevant published cyber security standards.}
\end{displayquote}
        
\begin{displayquote}
    \textit{(D4) The Code of Practice is supported by a mapping document and an open data JSON file that link each of the Code's guidelines against the main industry standards, recommendations and guidance. This mapping gives additional context to the Code's thirteen guidelines and helps industry to implement them.}
\end{displayquote}

\noindent D1 positions the guidelines as practical steps (which we interpret as: something to do, actions), but in D2 also as high-level outcomes (goals) to be reached by following the advice. D2 \label{foot:goodpractice} also mentions ``good practice'', but it is unclear whether this implies a \textit{practice} as we might define the term (see Table~\ref{tab:definitions} for explicit definitions), or just generally ``good things to do''.  

By our definitions (Table \ref{tab:definitions}), actions and outcomes are quite different. 
By D2, the guidelines intend to allow advice targets to select methods appropriate for their devices, suggesting the document should be used as a high-level set of outcomes to strive for rather than steps to follow. Readers may be misled about whether the guidelines alone suffice to reach security goals, or simply provide high-level statements of what should be done with external sources to provide further details (as apparently intended by D3--D4).

\begingroup
\renewcommand{\arraystretch}{1.5}

\begin{table}
    \centering
    \caption{Definitions of a few basic terms as we use them.}
    \label{tab:definitions}
    \vspace{-10pt}
    \footnotesize{
        \begin{tabular}{@{}p{0.85cm}p{7.25cm}@{}}
            \toprule
            Term & Definition \\ \midrule
            
            \textit{Outcome} & 
            A result of some prior activity; in our context, often the end goal of advice.\\
            
            \textit{Practice} & 
            A specific means or method, intended to achieve a given desired outcome.\\
            
            \textit{Action} & 
            An activity of one or more steps or a specific method, executed by a person or computer; in our context, often to achieve a desired outcome.\\
            
            \multirow{2}{2cm}{\textit{Actionable}\\\textit{Practice}} & 
            A practice (thus implying an action or unambiguous sequence of steps), whose means of execution is understood by targeted advice recipients.\\
            \bottomrule
        \end{tabular}
    }
\end{table}

\endgroup
        
Now moving to the \textit{ETSI Provisions}, they provide the following positioning statements \cite{ETSI2020}:

\begin{displayquote}
    \textit{(E1) The present document brings together widely considered good practice in security for Internet-connected consumer devices in a set of high-level outcome-focused provisions. The objective of the present document is to support all parties involved in the development and manufacturing of consumer IoT with guidance on securing their products.}
\end{displayquote}
        
\begin{displayquote}
    \textit{(E2) The provisions are primarily outcome-focused, rather than prescriptive, giving organizations the flexibility to innovate and implement security solutions appropriate for their products.}
\end{displayquote}
        
\begin{displayquote}
    \textit{(E3) [...] the focus is on the technical controls and organizational policies that matter most in addressing the most significant and widespread security shortcomings.}
\end{displayquote}
        
\begin{displayquote}
    \textit{(E4) Overall, a baseline level of security is considered; this is intended to protect against elementary attacks on fundamental design weaknesses (such as the use of easily guessable passwords).}
\end{displayquote}
        
\begin{displayquote}
    \textit{(E5) The present document provides a set of baseline provisions applicable to all consumer IoT devices. It is intended to be complemented by other standards defining more specific provisions and fully testable and/or verifiable requirements for specific devices which, together with the present document, will facilitate the development of assurance schemes.}
\end{displayquote}
        
\noindent E1 and E2 suggest the advice is intended to be high-level and focused on outcomes;
but E1 also positions these as ``good practice'', which here again (as earlier), is a term whose meaning is unclear. E3 mentions that the focus of the advice is technical controls and organizational policies. The former suggests to us a substantial level of detail (compared to the DCMS guidelines), but less specific than technical specifications (for which E5 defers to supplementary external documents, as do the DCMS guidelines). 

E4 positions the advice as protecting against ``elementary'' attacks, suggesting that the provisions are meant to address generic threats applicable to broad classes of IoT devices, versus advanced threats or those targeting specific environments (or users) or use cases. Recall that the DCMS 13 Guidelines are arranged in priority order by advice believed to be the most impactful. 

The term \textit{provision} itself is not explicitly defined in the ETSI Provisions, but based on context and our understanding of how it is used in the document, 
our informed guess is that this term is \textit{intended} to align with what we call a practice (Table \ref{tab:definitions}), as the provisions appear to be something that advice recipients are asked to execute to improve security. Despite our belief of this intent, in many cases the specified provisions lack ``specific means'' (as required by our explicit definition) to reach a goal.

As a summary regarding positioning (from our informal analysis), by our reading the DCMS and ETSI documents at first appear to be positioned similarly as high-level advice to be supported by more detailed advice, but on closer inspection the ETSI document is often clearer on how advice can be carried out. Both help in understanding the security landscape; the DCMS guidelines provide a high-level understanding of the general direction for security protections, while the ETSI document provides advice closer to our concept of practices for how to reach security goals. 

\subsection{Reference to External Advice}
\label{sec:criticism/externaladvice}
In our view, both documents fail to provide adequate direct reference to complimentary security advice as next-level information to support their high-level guidance, despite mentioning that their advice should be paired with such advice from external sources. We now give support for this view. 
    
The DCMS 13 Guidelines document suggests (in statements D2, D3, and D4 above) that stakeholders should use advice in the DCMS 1013-item dataset (mapped to each guideline in the mapping document) for technical details to follow the guidelines. The DCMS 13 document itself, however, does not clearly indicate \textit{which} advice items to use within the 1013-item dataset, and apparently expects stakeholders to sort this out for themselves. 

However, our previous work \cite{Barrera2022} suggests a relatively small proportion of items in the 1013-item dataset are actionable on their own; to follow a guideline, the DCMS 13 document implicitly expects readers (advice recipients) to have sufficient knowledge to find appropriate advice items from within the large set.  Whether the target audience actually has such knowledge has not, to our knowledge, been directly tested (this would in our opinion be informative).
    
The ETSI Provisions similarly do not provide specific implementation details for many provisions. The document does have a references section listing external sources, which some individual provisions reference. In most cases, the references are to relevant industry organizations, or entire documents (not specific pages or sections) from which readers would presumably independently locate and extract relevant information. Like the DCMS 13 Guidelines document, the ETSI Provisions suggests use of the 1013-item advice dataset \cite{IoTSecMap} (among others) to provide next-level advice. This raises the same question as above, about specifically which next-level advice to use from within the 1013-item dataset. 
    
As both documents suggest their guidelines and categories deliver \textit{good practice}, e.g, from the 1013-item advice dataset, it is unclear what exactly is expected of advice recipients when the documents do not actually specify (actionable) practices. Can an IT employee at an IoT manufacturer reliably select a small number of relevant items from the 1013-item dataset and properly execute the advice? Advice recipients are apparently required (expected) to select next-level advice from a suitable dataset or document (that they decide is appropriate) in order to reach desired security goals. 

\subsection{Target Audience}
\label{sec:criticism/targetaudience}
The DCMS 13 Guidelines document targets device manufacturers, IoT service providers, mobile app developers, and retailers \cite{DCMScodeOfPractice}; given an audience this broad, it is unclear what expertise advice recipients are expected to have. The document explicitly labels each guideline with one of these four targets (e.g., the first DCMS guideline ``\textit{no default passwords}'' is labelled as applying to device manufacturers). Thus not all guidelines apply to all four targets. Specific subgroups within each target (e.g., departments) are not described. 

In the ETSI document, the audience is described as ``\textit{organizations involved in the development and manufacturing of consumer IoT}'', much less specific than the DCMS audience; whether or not an evolution of the DCMS document, in this sense it does not improve.

We expect that following security advice would be the responsibility of relatively obvious subgroups, e.g., within manufacturer organizations (e.g., IT personnel in software development, or security positions); the DCMS document specifies the general targets, and we infer more specific audiences. While we are critical of the significantly more broad target description given in the ETSI document (and suggest the audience should be specified), given the document's appearance as an evolution of the DCMS document, we believe they are targeting similar audiences. 

A question to consider is: Within technically proficient target groups, do these documents assume that those who will follow security advice will be, e.g., security experts (with extensive knowledge and experience in security), or more typical developers and IT specialists familiar with basic security but not themselves security experts? This is relevant, as both documents expect the reader to consult external sources for implementation details, the comprehension of which may depend on one's security expertise. 
As we believe that it is beneficial for security advice to be actionable for the target audience, we argue that crafting advice specifically for  target audiences is important. We add this to our list of desirable characteristics for advice, and return to it in our systematic analysis.
    
\subsection{Distinct Advice Topics (fine-grained nature)}
\label{sec:criticism/improvements/subtopics}
Within an individual DCMS guideline, numerous sub-topics are sometimes offered. Consider DCMS-3: \textit{Keep software updated} \cite{DCMScodeOfPractice}:
    
\begin{displayquote}
    \textit{{\rm \{1\}} Software components in internet-connected devices should be securely updateable. {\rm \{2\}} Updates shall be timely and {\rm \{3\}} should not impact on the functioning of the device. {\rm \{4\}} An end-of-life policy shall be published for end-point devices which explicitly states the minimum length of time for which a device will receive software updates and the reasons for the length of the support period. The need for each update should be made clear to consumers and an update should be easy to implement. {\rm \{5\}} For constrained devices that cannot physically be updated, the product should be isolatable and replaceable.}
\end{displayquote}
    
\noindent This guideline has five distinct sub-topics (we inserted numbers \{$i$\} for exposition): {\rm \{1\}} secure updates, {\rm \{2\}} timely updates, {\rm \{3\}} updates that do not interrupt device function, {\rm \{4\}} published policy about update status, and {\rm \{5\}} ability to isolate and replace a device if it can not be updated. Each sub-topic might have distinct practices associated with it, but  the sub-topics are combined in one guideline. (The coding tree method used in our systematic analysis tags this guideline as \textit{Not Useful} (\textit{M1}), with a supplementary tag of \textit{Unfocused}.)
    
In contrast, the ETSI document offers individual provisions that each have a distinct topic. For example, consider Provision 3-1 under the same DCMS-3 heading \cite{ETSI2020}:
    
\begin{displayquote}
    ``\textit{All software components in consumer IoT devices should be securely updateable}''
    \quad [Notes and examples omitted]
\end{displayquote}
    
\noindent This provision is explicitly about secure updates, corresponding to {\rm \{1\}} above. Other sub-topics {\rm \{2\}}--{\rm \{5\}} are handled through separate provisions in the ETSI document. 
    
As a side note, given the 28 sub-topics in the DCMS 13 guidelines (this is the total number of sub-topics that we extract from the 13 guidelines in our systematic analysis), we might expect the ETSI document, if aiming to cover those same topics, to have about 28 provisions. However, the number of ETSI provisions (67) is more than twice this. The ETSI document uses IoT security advice documents found in the 1013-item dataset and additional reputable sources (noted in the ETSI Provisions introduction). 

It is unclear whether ETSI's increased number of advice items resulted from new topics being extracted from further sources, or if corresponding sub-topics found in the DCMS guidelines were more finely sliced. In either case, the ETSI Provisions have more specific, finer-grained items. Finally, that the major categories (the titles of each DCMS guideline) were largely retained while incorporating additional advice from further sources suggests that the original categories are largely representative of the general topics found to be important by IoT security advice-givers.
    
We believe it is reasonable to expect that advice recipients are able to extract sub-topics as we have above. Our systematic analysis below takes this into account by independently analyzing the DCMS guidelines from two perspectives: each guideline as presented in the source document, and with each sub-topic extracted and used as individual advice items (akin to ETSI's provisions). 
    
\subsection{Technical Content (actionability support)}
As discussed above, we view the positioning of both the DCMS 13 Guidelines and ETSI Provisions as somewhat self-contradictory, claiming to offer ``practical steps'' or ``technical controls'' (D1, E3), but also describing their advice as ``outcome-focused'' (D2, E2).
    
If the intention is to offer practical steps or technical controls, we would expect technical details included to provide advice targets clear instruction on how to execute the advice. If an advice item is vague about techniques expected to be used, the coding tree in our systematic analysis below is unlikely to   categorize the item as an actionable practice, and we expect that some advice recipients will be unclear about how to execute it without technical details from a further external source. The DCMS guidelines (and sub-topics therein) often briefly mention a technical approach without explicit or obvious implicit steps or actions to take. 
    
Consider DCMS-4 as example guideline lacking, in our view, sufficient technical details to enable reliable execution \cite{DCMScodeOfPractice}:
   
\begin{displayquote}
    \textit{Any credentials shall be stored securely in services and on devices. Hard-coded credentials in device software are not acceptable.}
\end{displayquote}

\noindent Not using hard-coded credentials would generally be regarded as a clear, requiring no further technical detail. In contrast is the  sentence requesting credentials be ``stored securely'' without explaining this (nor specific steps on how to achieve it). As technical details are largely absent from the DCMS 13 Guidelines, it appears the document (positioned ambiguously as discussed earlier) is intended as high-level advice with the expectation,
per D4 above, that external references be sought for technical detail on how to execute advice. 
    
ETSI provisions frequently suggest a technique or tool to use to achieve a goal, but often do not describe further execution detail. They often also include an example providing  context or describing a real-world scenario where application of the provision would benefit security. While such examples do not appear intended to replace technical detail, they add context that may help advice recipients reliably follow the advice. Like the DCMS 13 Guidelines, the apparent intent is that the advice is generally to be supplemented by external sources for technical details. 
    
\subsection{Summary of Informal Comparison} 
From our informal analysis, we find the DCMS 13 guidelines and ETSI provisions are positioned differently. Being especially interested in whether IoT security stakeholders can reliably execute provided advice, we note that the ETSI document appears to contain more technical content and detail than the DCMS guidelines (suggesting improved actionability).

\textbf{Positioning.} Both documents propose what are described as ``outcome-focused'' advice for consumer IoT devices, suggesting they aim to deliver high-level advice, but this is also arguably contradicted by positioning the advice as ``practical steps'' (DCMS) and ``technical controls'' (ETSI), setting up an expectation to deliver details enabling advice recipients to execute the advice.
    
\textbf{Reference to external advice.} Both documents rely on often vague references to external sources for further technical details of how to execute advice or reach security goals, leaving it unclear about which sources to follow to reach these goals. 
    
\textbf{Target audience.} Particularly in the ETSI Provisions, the target audience is vague, leaving unclear the required knowledge level of the audience expected to execute advice. The DCMS guidelines detail which of four general audiences each guideline targets. 
    
\textbf{Distinct advice topics.} Both documents categorize their advice in similar ways using high-level categories. However, the ETSI Provisions separate individual sub-items within each category into finer-grained stand-alone provisions, whereas the DCMS guidelines have multiple subtopics within a single block of text. 
    
\textbf{Technical content.} The DCMS 13 Guidelines provide little technical detail for how to follow advice, limiting its actionability. While the ETSI Provisions generally provide more technical detail, it appears that the intent of both documents is for advice recipients to reference external advice for further execution detail.

\section{Systematic Analysis}
\label{sec:experiment}
As opposed to the informal analysis above, we now carry out a systematic analysis of these same security advice documents, applying the SAcoding methodology \cite{Barrera2022} to these datasets. In particular, this allows a determination of the proportion of each advice set that is actionable (by our definition). This baseline measurement of actionability allows comparison of the DCMS 13 Guidelines and ETSI Provisions. From this, we proceed to other observations on whether the ETSI Provisions improve over the DCMS 13 Guidelines, with an eye towards more formally cross-checking the preliminary view of improvement from the informal analysis.

\subsection{Coding Tree Methodology (Overview)}
\label{sec:experiment/methodology}
\begin{figure}[bth]
    \centering
    \subfloat{{\includegraphics[width=0.35\textwidth]{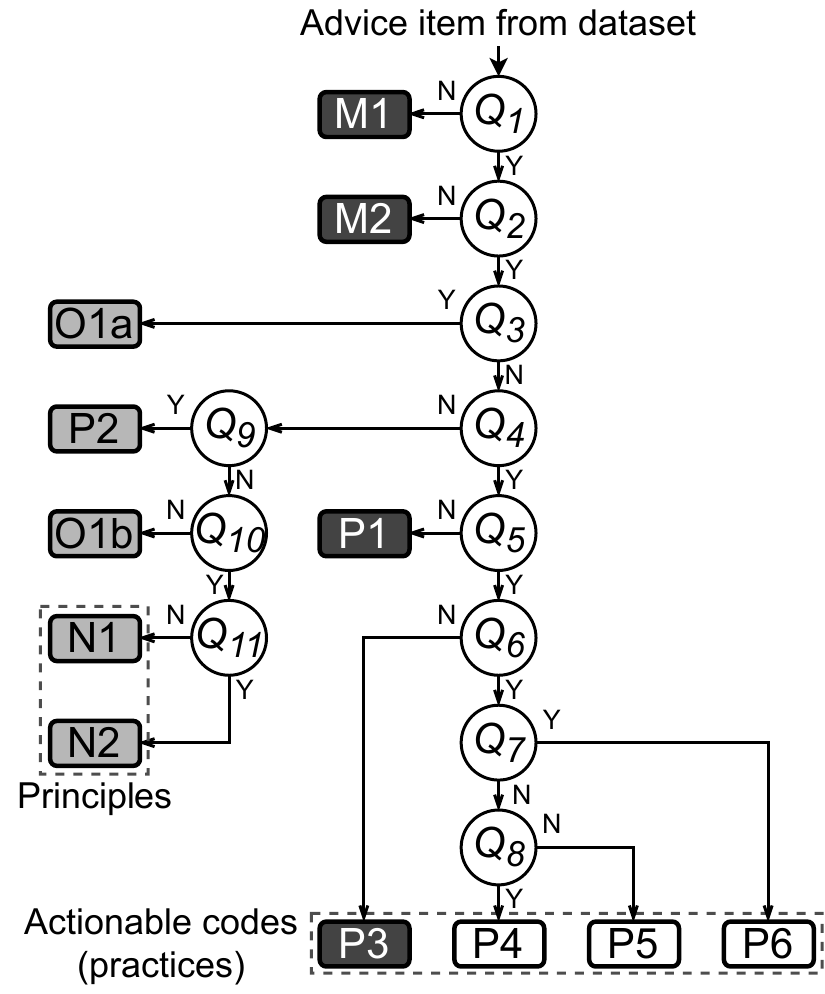}}}
    \vspace{-10pt}
    \caption{SAcoding tree \protect\cite{Barrera2022}. Answering questions $Q_i$ (Fig.~\ref{fig:treeQuestions}) leads to advice items being assigned leaf node codes (Fig.~\ref{box:tags}). Unshaded codes \textit{P4}--\textit{P6} are desirable (feasible) actionable practices; grey codes (\textit{O1a}, \textit{O1b}, \textit{N1}, \textit{N2}, \textit{P2}) are not actionable but may be useful in some contexts.  We view dark-shaded codes (\textit{M1}, \textit{M2}, \textit{P1}, \textit{P3}) as non-desirable targets for advice.
    }
    \label{fig:flowchart2}
\end{figure}

We summarize here our \textit{security advice coding} (SAcoding) methodology, previously introduced and used \cite{Barrera2022} on the 1013-item dataset of security advice items.\footnote{
    Our previous work \cite{Barrera2022} analyzed the V3 dataset \cite{IoTSecMap}; V4 is now available.
} It was designed using inductive coding \cite{Corbin2008}. Fig.~\ref{fig:flowchart2} shows the coding tree, which a human coder uses to assign codes to advice items by answering questions $Q_i$ given in Fig.~\ref{fig:treeQuestions}. The method does not analyze the quality of the technical content of advice, but rather classifies advice items into ``type'' categories, per the codes in Fig.~\ref{box:tags}.

\begin{figure}[bth]
    \centering
    \qquad
    \small{
        \parbox[t]{0.7\textwidth}{
            \newcommand{\Spacing}{-4pt}
            \vspace{\Spacing}
            \paragraph{Q1.} Is the item conveyed in unambiguous language, and relatively\\ focused?
            \vspace{\Spacing}
            \paragraph{Q2.} Is it arguably helpful for security?
            \vspace{\Spacing}
            \paragraph{Q3.} Is it focused more on a desired outcome than how to achieve it?
            \vspace{\Spacing}
            \paragraph{Q4.} Does it suggest a security technique, mechanism, software tool,\\ or specific rule?
            \vspace{\Spacing}
            \paragraph{Q5.} Does it describe or imply steps or explicit actions to take?
            \vspace{\Spacing}
            \paragraph{Q6.} Is it viable to accomplish with reasonable resources?
            \vspace{\Spacing}
            \paragraph{Q7.} Is it intended that the end-user carry this item out?
            \vspace{\Spacing}
            \paragraph{Q8.} Is it intended that a security expert carry this item out?
            \vspace{\Spacing}
            \paragraph{Q9.} Is it a general policy, general practice, or general procedure?
            \vspace{\Spacing}
            \paragraph{Q10.} Is it a broad approach or security property?
            \vspace{\Spacing}
            \paragraph{Q11.} Does it relate to a principle in the design?
        }
    }
    \vspace{-10pt}
    \caption{SAcoding tree questions \protect\cite{Barrera2022}.
    }
    \label{fig:treeQuestions}
\end{figure}

\begin{figure}[t]
    \noindent
    \fbox{\parbox[t]{0.48\textwidth}{
        \small{
            \vspace{-4pt}
            \paragraph{M1. Not Useful (too vague/unclear or multiple items):} Advice that does not make sense from a language perspective (e.g., unclear due to grammar), or is not focused on a specific task/action to complete.
            
            \paragraph{M2. Beyond Scope of Security:} Advice that is not clearly an item that might benefit security.
            
            \paragraph{N1. Security Principle:} Advice framed as a general rule that applies broadly, and has historically improved security outcomes or reduced exposures. 
            
            \paragraph{N2. Security Design Principle:} Advice that suggests a \textit{Security Principle}, and more specifically, applies to the design phase of a product's lifecycle.
            
            \paragraph{O1a/b. Desired Outcome:} Advice that suggests a generic, high-level end goal to be attained (without specifying a particular method by which to reach it).
            
            \paragraph{P1. Incompletely Specified Practice:} Advice that suggests a technical direction of a practice (e.g., a technical mechanism, specific rule), but lacking clear indication of steps to follow, and thereby considered ``non-actionable''.
            
            \paragraph{P2. General Practice or General Policy:} Advice that is not explicit about techniques or tools, but indicates a general approach; may be policy-related. Considered non-actionable (despite name) due to general, unspecific nature. 
            
            \paragraph{*P3. Infeasible Practice:} A practice that would require unreasonable resources (time, money), thereby failing most or all cost-benefit analyses.  
            
            \paragraph{*P4. Specific Practice---Security Expert:} A practice requiring a security expert to implement, e.g., in-depth knowledge or experience; often requires steps not clearly specified in advice to be inferred.
            
            \paragraph{*P5. Specific Practice---IT Specialist:} A practice that typical IT workers could execute, using basic (vs security expert) professional knowledge of security.
            
            \paragraph{*P6. Specific Practice---End-User:} A practice that typical end-users could execute, e.g., by directly interacting with a device, mobile app, or cloud service.
        }
    }}
    \vspace{-10pt}
    \caption{SAcoding tree codes, based on Barrera et al.\ \protect\cite{Barrera2022}. Codes preceded by an asterisk (*) are considered actionable.} 
    \label{box:tags}
\end{figure}

We use the SAcoding tree on three datasets: two derived from the DCMS 13 Guidelines, and one from the ETSI Provisions. For the first of the two from the DCMS document, for an actionability assessment of the guidelines as documented, we extract each guideline as a whole (the entire guideline as one item). For the second set, we extract all sub-topics from each of the 13 guidelines, allowing assessment of each sub-topic as an independent advice item. From the ETSI Provisions we extract each provision (including associated examples and notes) from each of the 13 categories plus the extra category noted earlier. Each ETSI provision is largely self-contained, typically focusing on one topic, and need not be subdivided. These extractions result in three datasets that we will compare: 
    
\begin{itemize}
    \item DCMS \textit{Full} guidelines (13 items)
    \item DCMS \textit{Sub-Topics} guidelines (28 items)
    \item ETSI \textit{Provisions} (67 items)
\end{itemize}
    
\noindent For each item in each dataset, using a software interface tool, one author applied the SAcoding method. This assigns a code (less formally, \textit{tag}) to each item. Each code is pre-classified as actionable or not (as designated in Fig.~\ref{box:tags}), allowing a count of actionable advice items in each dataset (and thus the proportion of each set that is actionable). Our use of this method matches our previous work \cite{Barrera2022}.

Herein, a lightweight cross-check was added with a second coder looking for inconsistencies, e.g., between codes assigned to DCMS Full and Sub-topics, and how codes $N1$, $N2$ (principles) were used. On discussion, this resulted in changing codes assigned to a few items. Inter-coder agreement on testsets between pairs of three coders was taken into account during the iterative design of SAcoding \cite{Barrera2022}. Separate work~\cite{Barrera2022a} compares results from two coders on a full 1013-item dataset.  

Regarding use of one coder vs.\ a full cross-check with a second coder, the present work aims to compare two advice documents. For this, we believe biases or subjective interpretations of questions or instructions by the single coder are likely to be consistent across their coding of both datasets, and thus expect potential inconsistencies between different coders, while undesirable, are perhaps less relevant.

\subsection{Results of coding}
\label{sec:experiment/results}
Recall that \textit{DCMS-i} and \textit{ETSI-i} indicate the same 13 category headers, conveying  the same or very similar information (e.g., DCMS ``\textit{Monitor system telemetry data}'' \cite{DCMScodeOfPractice} versus ETSI ``\textit{Examine system telemetry data}'' \cite{ETSI2020}); however the content of DCMS guidelines differs from that of ETSI provisions. We analyze the content. 
    
\begingroup
\setlength{\tabcolsep}{3pt} 

\begin{table}[t]
    \centering
    \caption{DCMS Sub-Topic and Full guideline coding results. \textit{n} is number of sub-topics manually extracted. \textit{Sub-Topics} are each assigned a code. \textit{Full} column gives code assigned to overall guideline. *denotes actionable codes of Fig.~\ref{box:tags}. }
    \vspace{-10pt}
    \label{tab:taggingresults/DCMS}
    \small{
        \begin{tabular}{@{}lccccccrr@{}}
            & & \multicolumn{5}{c}{Frequency of code} & \\
            & & \multicolumn{5}{c}{assigned to Sub-Topic} & Full\\
            \cmidrule(lr){3-7} \cmidrule(lr){8-8}
            Guideline & n & P1 & P2 & *P5 & N2 & O1a & \\
            \midrule
            DCMS-1  & 1 &  &  & 1 &    &  & *P5\\
            DCMS-2  & 2 & 1 &  & 1 &    &  & M1\\
            DCMS-3  & 5 & 1 &  & 1 &    & 3 & M1\\
            DCMS-4  & 2 & 1 &  & 1 &    &  & M1\\
            DCMS-5  & 2 & 1 &  &  &    & 1 & M1\\
            DCMS-6  & 1 &  &  &  &   1 &  & N2\\
            DCMS-7  & 2 & 1 &  & 1 &    &  & M1\\
            DCMS-8  & 3 & 1 &  & 2 &    &  & M1\\
            DCMS-9  & 4 & 1 &  &  &    & 3 & M1\\
            DCMS-10 & 1 & 1 &  &  &    &  & P1\\
            DCMS-11 & 2 & 1 & 1 &  &    &  & M1\\
            DCMS-12 & 2 & 1 & 1 &  &    &  & M1\\
            DCMS-13 & 1 & 1 &  &  &    &  & P1\\
            \midrule
            Total (28) &  & 11 & 2 & 7 &  1 & 7 \\
            \midrule
            Proportion of Total &  & 39.3\% & 7.1\% & 25.0\% &  3.6\% & 25.0\% \\
            \bottomrule
        \end{tabular}
    }
    \vspace{-10pt}
\end{table}

\endgroup

\label{page:DCMSresults}
Table~\ref{tab:taggingresults/DCMS} shows the results for tagging both the DCMS Sub-Topics set and the DCMS Full set. Only tags represented in the analysis results are listed in the table; unused tags are omitted. For the Full guideline tagging, only 1 of 13 guidelines (DCMS-1) was assigned an actionable tag (here, \textit{P5}). For Sub-Topics, 7 of 28 sub-topic items (25\%) were tagged \textit{P5} (the only actionable tag that appeared in the results); the remaining 75\% are non-actionable tags. Note that only 6 of the 11 tags available (from Fig.~\ref{box:tags}) appear in the Table~\ref{tab:taggingresults/DCMS} results.  
    
\begingroup
\setlength{\tabcolsep}{3pt} 

\begin{table}[b]
    \centering
    \caption{ETSI Provisions coding results. $n$ is number of provisions in category. Other column headings denote codes assigned using coding tree. *actionable codes from Fig.~\ref{box:tags}.}
    \vspace{-10pt}
    \label{tab:taggingresults/ETSI}
    \small{
        \begin{tabular}{@{}lcccccccc@{}}
                        & n & P1    & P2& *P4 & *P5 & O1a & N2 & M1 \\
            \midrule
            ETSI-1     & 5 & 2     &   &   & 3 &   &   & \\
            ETSI-2     & 3 &       & 1 &   & 2 &   &   &  \\
            ETSI-3     & 16 & 5     & 4 &   & 5 & 1 &   & 1 \\
            ETSI-4     & 4 &       &   & 1 & 3 &   &   &  \\
            ETSI-5     & 8 & 5     & 1 &   & 1  &   & 1  &  \\
            ETSI-6     & 9 & 2     & 1 &   & 3  & 2  & 1  &  \\
            ETSI-7     & 2 & 1     &   &   & 1  &   &   &  \\
            ETSI-8     & 3 & 2     &   &   & 1  &   &   &  \\
            ETSI-9     & 3 &       &   &   & 1  & 2  &   &  \\
            ETSI-10     & 1 & 1     &   &   &   &   &   &  \\
            ETSI-11     & 4 & 2     &   &   & 2  &   &   &  \\
            ETSI-12     & 3 & 1     &   &   & 2  &   &   &  \\
            ETSI-13     & 1 & 1     &   &   &   &   &   &  \\
            ETSI-DP     & 5 & 1     &   &   & 4  &   &   & \\ 
            \midrule
            Total (67) & &  23 & 7 & 1 & 28 & 5 & 2 & 1\\
            \midrule
            Proportion of Total & & 34.3\% & 10.4\% & 1.5\% & 41.8\% & 7.5\% & 3.0\% & 1.5\% \\
            \bottomrule
        \end{tabular}
    }
    \vspace{-10pt}
\end{table}

\endgroup  
    
\begin{figure}[t]
    \centering
    \includegraphics[width=0.42\textwidth]{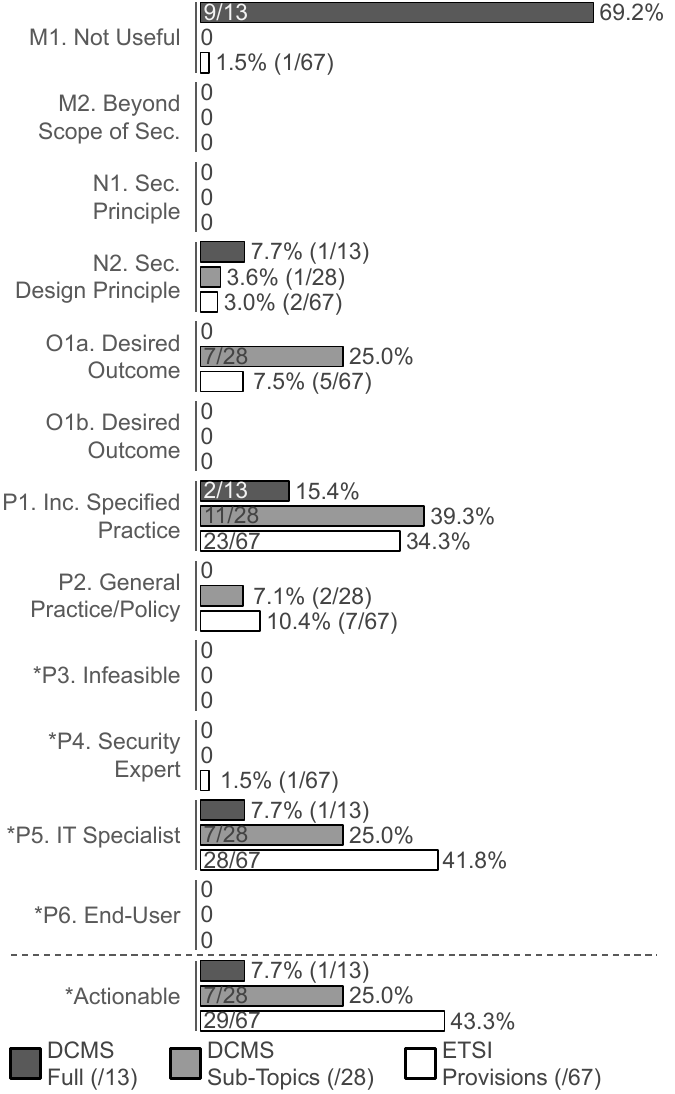}
    \vspace{-8pt}
    \caption{Tag distribution across advice datasets. Bars depict proportion of items in each dataset tagged with specified code. \textit{Actionable} bars depict proportion of each dataset assigned actionable codes. Based on data from Tables \ref{tab:taggingresults/DCMS} and \ref{tab:taggingresults/ETSI}. *indicates actionable codes. Shading follows Fig.~\ref{fig:flowchart2}'s scheme. 
    }
    \label{fig:setsbars}
\end{figure}
\raggedbottom
    
Table~\ref{tab:taggingresults/ETSI} shows the results of tagging the ETSI Provisions, with 7 tags appearing across its 67 provisions (those in Table~\ref{tab:taggingresults/DCMS} plus \textit{P4}); absent are \textit{P3}, \textit{P6}, \textit{M2}, \textit{O1b}. A combined 43\% of provisions were tagged with actionable tags (here including \textit{P4} and \textit{P5}). 
    
Figure~\ref{fig:setsbars} shows the proportion of advice from the three datasets (DCMS \textit{Full} and \textit{Sub-Topics}, and  ETSI \textit{Provisions}) assigned each code. We  give our interpretation of these results next. 

\subsection{Interpretation and Comparative Analysis}
\label{sec:experiment/discussion}
Based on our methodology, the ETSI Provisions improve over the DCMS 13 Guidelines in terms of proportion of actionable advice. 

As a first observation, the \textit{Incompletely Specified Practice} (\textit{P1}) code was used 15.4\%, 39.3\%, and 34.3\% of the time for the DCMS Full, DCMS Sub-Topics, and ETSI Provisions sets, respectively (Fig.~\ref{fig:setsbars}). The frequency with which code P1 was assigned (through the coding tree) signals to us a heavy reliance on external advice; due to a lack of technical detail within advice items themselves, and lack a direct reference to external sources for such detail, these items are not actionable by our definition. 

\label{page:misinterpretation} The DCMS Full results show significantly fewer \textit{P1} codes (which intuitively appears desirable) than the DCMS Sub-Topics and ETSI Provisions, but this is perhaps misleading, as in 9 of 13 cases (Table~\ref{tab:taggingresults/DCMS}) the Full guidelines did not have an opportunity to reach \textit{P1} in the coding tree due to being immediately assigned \textit{M1}. (From Fig. \ref{fig:flowchart2}, note that $Q_1$ asks whether an item is relatively focused; many Full guidelines yield a \textit{no} answer, resulting in tag \textit{M1} being assigned.)

\label{page:unusedcodes}     
Fig.~\ref{fig:setsbars} shows that tags \textit{M2} (\textit{Beyond the Scope of Security})  and \textit{P3} (\textit{Infeasible Practice}) were not assigned to items in any of our three datasets. This matches our previous analysis \cite{Barrera2022}, presumably for the same reason---when advice in that 1013-item dataset was tagged actionable, it was almost always feasible to carry out (cf.\ $Q_6$); and advice intended as security advice was rarely assigned \textit{M2} at $Q_2$.

\textit{P4} (\textit{Security expert}) appeared just once in the ETSI Provisions analysis, and \textit{P6} (\textit{End-User}) 0 times. This suggests both advice documents matched their asserted target audience (manufacturers and other pre-deployment stakeholders).
    
Fig.~\ref{fig:setsbars} also shows that in terms of the proportion of dataset items that are actionable, the ETSI Provisions (at 43.3\%) substantially improve over the DCMS Full (7.7\%) and Sub-Topics (25.0\%) guidelines. While we had expected that all three datasets would be less actionable than the next-level 1013-item dataset that they reference (found to be 32--33\% actionable in our previous analysis \cite{Barrera2022}), Fig.~\ref{fig:setsbars} shows our analysis found one of our three, the ETSI Provisions, is more actionable. We attribute this improvement to the ETSI Provisions having been refined into a finer-grained, focused set of 67 items. We next discuss the results of each set tagging.
    
\subsubsection{DCMS 13 guidelines (tagging results and actionability)}
It is unsurprising that the DCMS full guidelines considered in their entirety (containing more than one topic) are so frequently tagged \textit{M1} (\textit{Not Useful---too vague/unclear or multiple items}). The coding tree methodology's $Q_1$ was designed to identify vague and unclear items (signalling these as candidates for clarification); beyond this, our work indicates that the tree can provide more insightful output if such unfocused advice items are manually separated by preprocessing into finer-grained advice items. If producing actionable security advice is the goal of an advice giver, the advice items in source documents themselves likewise should be refined into narrower individual items dwelling on single topics rather than spanning several topics. This is illustrated in Table~\ref{tab:taggingresults/DCMS} and Fig.~\ref{fig:setsbars}. 
        
Our tagging of the DCMS guideline Sub-Topics provided a more granular look at the advice. While our view is that the guidelines were intended to be digested as a whole (based on being formally presented as a paragraph and within a highlighted section as noted), our results shown in Fig.~\ref{fig:setsbars} (with higher actionability, when sub-topics are rated individually than when grouped) indicate that it may be more useful to present each sub-topic as a distinct piece of advice---that is, if actionability is important to the advice giver. In fact, this is the direction taken by the ETSI Provisions.
        
While Table~\ref{tab:taggingresults/DCMS} shows \textit{M1} as the most common code by far in the Full dataset, it did not appear in the Sub-Topics set results.  There, sub-topics were extracted and apparently all items were understandable and contained only one topic (otherwise $Q_1$ would trigger a \textit{no}, yielding \textit{M1}). All \textit{M1} codes in the Full set were accompanied by the \textit{Unfocused} supplementary topic being selected by the coder.

Considering the sub-topics individually, 25\% were categorized as actionable (Table~\ref{tab:taggingresults/DCMS}). This is an increase from 8\% for the Full guidelines (1 of 13 guidelines was tagged actionable). As a side point, in both cases, some items were actionable by IT specialists (corresponding to code \textit{P5}, occurring proportionally as a ratio 1/13 and 7/28 for Full and Sub-Topics sets, respectively), but none were tagged as \textit{P4} (requiring security experts). This suggests that the sub-topics that \textit{are} actionable were appropriate for IT specialists, but not requiring the further expertise of security experts. While not dominant in number, we view the occurrences of \textit{P5} (versus \textit{P4}) positively, as we expect IoT device manufacturers to typically have IT specialists as developers. Thus, the subset of Sub-Topics advice that is actionable matches what we interpret as the target audience. 
        
\subsubsection{ETSI Provisions (tagging results and actionability)}
For the ETSI tagging, Fig.~\ref{fig:setsbars} shows that 42\% of advice  was tagged as \textit{P5} (\textit{Specific Practice---IT Specialist}). One practice (1/67), in ETSI-4 of Table~\ref{tab:taggingresults/ETSI}, was tagged as requiring a Security Expert (\textit{P4}). Of actionable codes, 28/29 are \textit{P5}, matching what we believe is the target audience (IT specialists rather than security experts). A greater proportion of the ETSI Provisions set were tagged as \textit{P5} than in the DCMS Sub-Topics set (41.8\% vs. 25\%), implying the ETSI provisions are more appropriate than the DCMS sub-topics for what we believe is the ideal target audience for the advice. Combined, actionable codes make up 43.3\% of ETSI Provisions. 
        
Comparing the ETSI Provisions advice set to the DCMS Full and Sub-Topics sets, the ETSI Provisions set is significantly more actionable, as shown by  the Actionable bars in Fig.~\ref{fig:setsbars}. 

\subsubsection{Comparative improvement} 
We now summarize the results from our systematic comparative analysis of the two main advice documents, supporting several aspects from the initial informal analysis that suggest the  ETSI Provisions improve over the DCMS 13 Guidelines.
    
\textbf{(1) Actionability of advice.}
The ETSI Provisions are considerably more actionable than the DCMS Sub-Topics and single guidelines. The actionability of the DCMS Sub-Topics (25\%) more than triples the corresponding 7.7\% for the single guidelines, but is itself almost doubled by ETSI Provisions (43\% actionable). We attribute this largely to the improvement in technical detail within the ETSI Provisions (next item). 
        
\textbf{(2) Technical detail included more often.}
For the DCMS Sub-Topics set, a combined 64\% (18 of 28) evoked a \textit{yes} at $Q_4$ (Fig.~\ref{fig:flowchart2}), suggesting they described a security technique, mechanism, software tool, or specific rule. This is the coding tree's first technical branch toward an actionable code. 11 of these 18 items exited the path toward an actionable code at $Q_5$ (which asks: does the advice item describe or imply steps to take), moving instead to \textit{P1} (\textit{Incompletely Specified Practice}). In the case of the DCMS Sub-Topics set, this was typically because of a lack of technical detail in the advice. In contrast, for the ETSI Provisions, 78\% of the 67 items progressed to an actionability branch at $Q_4$, and 43\% likewise at $Q_5$. We interpret the proportion of a dataset that is actionable as a signal of how technically detailed the dataset is---an advice item reaching $Q_6$ (beyond which all tags are actionable) has enough technical detail to be considered actionable. Thus, we see significant improvement in the level of technical detail in the ETSI Provisions set (43.3\% of provisions evoked a \textit{yes} at $Q_5$, versus 25\% of DCMS sub-topics). 
        
\textbf{(3) Fewer incomplete practices.}
For the DCMS Sub-Topics set, 39.3\% were tagged \textit{P1} (\textit{Incompletely Specified Practice}). This drops to 34.3\% for the ETSI Provisions, suggesting that when these provisions are able to make it to $Q_5$ about technical steps, they continue on the actionable path by specifying technical steps more frequently than the DCMS Sub-Topics. (As evident from the tree, a \textit{yes} at $Q_5$ branches to actionable codes (\textit{P3}--\textit{P6}), while answering \textit{no} yields the \textit{P1} leaf.) While only 15.4\% of the guidelines from the DCMS Full set were tagged \textit{P1} (significantly fewer incomplete practices than the other two sets), as discussed earlier, this is explained by the early assignment of tag \textit{M1} rather than a specific improvement in the DCMS Full advice dataset. \\
    
In summary, facilitated by the coding tree methodology, we analyzed two security advice documents from authoritative sources. We found that the ETSI Provisions improve on the DCMS 13 Guidelines in proportion of actionable advice, frequency of technical detail, and a reduction in the number of incomplete practices. 

\section{Related Work}
Literature on security advice has included discussion of \textit{usability}, but most commonly related to end-users (and often in the Internet of Computers), and less often exploring impact on IT workers---which is our focus, including advice for pre-deployment stakeholders and a focus on IoT. We note here a small sampling of related work on security advice for both IT workers and end-users. For related work on security best practices and technical issues related to IoT practices, see Barrera et al.\ \cite{Barrera2022}.  

Redmiles et al.\ \cite{Redmiles2018} measure the readability of security advice from both expert and non-expert online sources. In our work, readability is also crucial; the coding tree disqualifies vague or unfocused advice at the first branch of the tree. Other work by Redmiles et al.\ \cite{Redmiles2020} collected and analyzed a large set of end-user security advice and found most advice was perceived as actionable, but users were unclear about which advice is most important to execute. Mannan and van Oorschot \cite{Mannan2008} reviewed security advice provided by major banks to provide secure access to online banking websites, and found much of the advice too difficult for online banking users to follow. Herley \cite{Herley2009} suggests that end-users commonly reject security advice because following it brings greater costs than benefits. 

DeKoven et al.\ \cite{Dekoven2022} relate laptop and desktop end-users (vs.\ IoT or mobile phone users) following common security advice, to security outcomes, i.e., whether hosts end up compromised; they note the lack of strong correlation between compromise and not following recommended advice.  In contrast, the SAcoding method used herein primarily involves categorizing advice intended for product manufacturers, without attempting to analyze its content technically, or correlating the following of advice to security outcomes.

Complementary to security advice, Stevens et al.\ \cite{Stevens2020} investigate compliance with three security standards. They find that security concerns may remain even when standards are conformed to, suggesting the reviewed standards (cf.\ security practices compared herein) may be improved to reduce security issues. The stipulation of security advice (i.e., some authoritative entity requiring it, e.g., to comply with policy) is also discussed by Barrera et al.\ \cite{Barrera2022}. 

Acar et al.\ \cite{Acar2017} note that software developers often find security advice lacking (out of date, missing concrete examples or important topics), and find technical detail or pointers to external advice frequently lacking in advice from major industry organizations. Renaud \cite{Renaud2016} notes in small and medium-sized enterprises, large volumes of security advice lead to uncertainty and confusion; and advocates for clearer, fewer, and simpler sets of security advice. Assal and Chiasson \cite{Assal2018} note that available information resources for software development security practices vary in level of technical detail; and report frequent non-compliance with software development security advice, e.g., circumventing annoying or frustrating security features that hinder development speed. RFC 2119 \cite{Bradner1997} defines terminology (e.g., MUST, SHALL, SHOULD) defining how to interpret RFCs; while not directly about security advice, such well-defined (and well-adopted) terms clarify the expectation on advice followers. This is but a subset of papers on usability and utility of general security advice, both for end-users \cite{Redmiles2020, Herley2009, Florencio2007b} and IT workers \cite{Assal2018, Acar2017, Renaud2016}. 

While our SAcoding method \cite{Barrera2022} used herein does not code qualitative data produced by users \cite{Corbin2008}, the DCMS and ETSI advice datasets coded herein are comprised of qualitative data (security advice items). Many others have used qualitative coding methodologies for security-related topics; we mention here just a few. Huaman et al.\ \cite{Huaman2021} coded user feedback about password managers, such as issues with how password managers interact with websites (e.g., difficulty detecting input fields, website JavaScript preventing automated input). Naiakshina et al.\ \cite{Naiakshina2017} coded interview responses on how participants used secure password storage mechanisms; they found that participants prioritized web application functionality ahead of security, and that security advice for  password storage was not being followed by participants. Krombholz et al.\ \cite{Krombholz2017} coded verbalized participant thoughts during a system configuration task and found configuring TLS for a web server was too complex for participants, and suggested that default server configurations should be stronger. 

\section{Concluding Remarks}
Given IoT's popularity and rapid growth over the past two decades, many organizations have offered security advice to IoT security stakeholders. While the advice appears well-intentioned, it has been less clear how actionable that advice is for these stakeholders. We argue that actionable advice supports good security development practices, which have been found lacking in consumer IoT \cite{Alrawi2019}. 

Our analysis of the DCMS 13 Guidelines and ETSI Provisions suggests that constructing sets of actionable security advice may not be as simple as the wide breadth of existing advice would make it seem. What is positioned to be widely ``appropriate'' advice may be ineffective for mismatched audiences (e.g., if expert level security advice is used by non-experts). Despite positive aspects, our analysis found both these documents leave room for improvement.

Using the SAcoding method, we found an improvement in actionability of IoT security advice from the UK DCMS 13 Guidelines to the ETSI Provisions. Such analysis provides a basis from which to begin measuring the quality of security advice, including characteristics beyond actionability. We encourage systematic assessment of other important elements of security advice documents and datasets. It would be interesting in follow-on work to recruit as additional coders, industry experts and security practitioners, and compare their interpretations of both advice documents with our interpretations and results herein. 

Our analysis suggests the following aspects be considered in creating security advice: target audiences, level of technical detail, references to external advice, granularity of advice items, and positioning. Our concluding thoughts on these follow. 

\textbf{Explicit declaration of target audience.} An explicit declaration and characterization of  target audiences of security advice is important in our view. This plays directly into whether advice is actionable, as advice is often tailored for a specific audience and their knowledge level. Without a declaration, unintended audiences who use the advice may struggle to understand what is expected, or may lack the knowledge to successfully (reliably) execute advice.

\textbf{Appropriate level of technical details.} Related to suiting target audiences,
advice documents should include a level of detail matching their positioning. In our view, that level should be sufficient to provide target recipients (implicitly or explicitly) unambiguous and clear steps for how to reach desired outcomes. This may be accomplished by advice itself containing details (such that an advice recipient can understand, e.g., the technical mechanisms suggested), or referencing external sources for next-level details (see next item). We note that the concepts of actionable practices and target audiences are embedded in the SAcoding method. 
    
\textbf{Explicit pointers to next-level detail.} Related to the previous item, if advice itself lacks sufficient technical detail to be directly executed by a target audience, it should specifically reference resources containing next-level details. This is a good general rule beyond IoT security advice---for example, applying also to security toolkit error messages for software developers, such as technical documentation related to certificate validation errors \cite{Ukrop2019}. Related to this, we note that the likelihood of reliably executing advice diminishes if referenced documents themselves send advice recipients down a chain of further references, e.g., with critical details three levels deeper. Aside from overburdening advice recipients, later documents may also be written for different target audiences. 

\textbf{Fine-grained advice items.} It appears beneficial to package advice as fine-grained items addressing narrower or single topics. Our systematic analysis comparing the DCMS Full and DCMS Sub-Topics datasets supports this---when blocks of advice text from the Full set were split into extracted sub-items for the Sub-Topics set, all \textit{Not Useful} (\textit{M1}) codes were replaced by codes conveying more meaningful information reflecting sub-item content. Analysis of the ETSI Provisions, which are finer-grained than the DCMS Full set, also supported this view; the provisions attracted in total just one \textit{Not Useful} code among 67 items.

\textbf{Positioning.} We suggest that security advice documents clearly indicate whether their advice is intended to convey, e.g., practices, guidelines, or requirements; this may help avoid contradictions such as documents suggesting that advice that specifies only end-goals is also actionable. As discussed elsewhere \cite{Barrera2022}, terminology used to convey security advice is important, and may signal how it is designed, or expected to be used. If we agree to associate definitions of the term \textit{practice} with actionability, then documents accurately advertising practices (or \textit{best practices}) are more likely to be reliably followed than those suggesting only desired outcomes. 

We believe that more work is needed on both means to create reliably executable security advice, and on means to evaluate it; the two are clearly related. Better tools are needed to help those who provide guidance (advice-givers) to craft advice that can clearly and coherently guide advice recipients. As steps in this direction, we encourage use and refinement of new systematic methods that aim to measure or grade security advice. We expect that any advances improving security advice for IoT will naturally apply also to the Internet of computers, and vice versa.

\begin{acks}
    We thank anonymous referees for helpful comments. The second author is Canada Research Chair in Authentication and Computer Security, and acknowledges NSERC for funding the chair and a Discovery Grant. 
\end{acks}

\bibliographystyle{ACM-Reference-Format}
\bibliography{bib}

\appendix

\section{Systematic Analysis Coding Data}
Tables~\ref{tab:taggingoutput/ETSIProvision} and \ref{tab:taggingoutput/DCMSSubTopic} resp.\ show our coding for the ETSI Provisions and both DCMS datasets (Full and Sub-Topics). These are included to allow comparison by others who may wish to carry out their own independent coding using SAcoding or any other method.

\begingroup
\setlength{\tabcolsep}{3pt}
\renewcommand{\arraystretch}{1.0}

\begin{table}[H]
    \centering
    \caption{Our coding of ETSI Provisions dataset (67 items), using the SAcoding method. Table \ref{tab:taggingresults/ETSI} uses this data.}
    \label{tab:taggingoutput/ETSIProvision}
    \small{        
        \begin{minipage}{0.15\textwidth}
        \centering
            \begin{tabular}{ll}
                \toprule
                Provision & Code \\
                \midrule
                1-1 & P5 \\
                1-2 & P1 \\
                1-3 & P1 \\
                1-4 & P5 \\
                1-5 & P5 \\
                \midrule
                2-1 & P5 \\
                2-2 & P2 \\
                2-3 & P5 \\
                \midrule
                3-1 & O1a \\
                3-2 & P1 \\
                3-3 & P1 \\
                3-4 & M1 \\
                3-5 & P1 \\
                3-6 & P5 \\
                3-7 & P1 \\
                3-8 & P2 \\
                3-9 & P1 \\
                3-10 & P5 \\
                3-11 & P5 \\
                3-12 & P5 \\
                3-13 & P2 \\
                3-14 & P2 \\
                3-15 & P2 \\
                3-16 & P5 \\
                \bottomrule
            \end{tabular}
        \end{minipage}
        \begin{minipage}{0.15\textwidth}
        \centering
            \begin{tabular}{ll}
                \toprule
                Provision & Code \\
                \midrule
                4-1 & P5 \\
                4-2 & P4 \\
                4-3 & P5 \\
                4-4 & P5 \\
                \midrule
                5-1 & P1 \\
                5-2 & N2 \\
                5-3 & P1 \\
                5-4 & P5 \\
                5-5 & P2 \\
                5-6 & P1 \\
                5-7 & P1 \\
                5-8 & P1\\
                \midrule
                6-1 & P1 \\
                6-2 & O1a \\
                6-3 & O1a \\
                6-4 & P5 \\
                6-5 & P5 \\
                6-6 & P5 \\
                6-7 & N2 \\
                6-8 & P1 \\
                6-9 & P2 \\
                \midrule
                7-1 & P1 \\
                7-2 & P5 \\
                \bottomrule
                $~~$ \\
            \end{tabular}
        \end{minipage}
        \begin{minipage}[H]{0.15\textwidth}\centering
            \begin{tabular}{ll}
                \toprule
                Provision & Code \\
                \midrule
                8-1 & P1 \\
                8-2 & P1 \\
                8-3 & P5 \\
                \midrule
                9-1 & O1a \\
                9-2 & O1a \\
                9-3 & P5 \\
                \midrule
                10-1 & P1 \\
                \midrule
                11-1 & P1 \\
                11-2 & P1 \\
                11-3 & P5 \\
                11-4 & P5 \\
                \midrule
                12-1 & P1 \\
                12-2 & P5 \\
                12-3 & P5 \\
                \midrule
                13-1 & P1 \\
                \midrule
                DP-1 & P5 \\
                DP-2 & P5 \\
                DP-3 & P5 \\
                DP-4 & P1 \\
                DP-5 & P5 \\
                \bottomrule \\
                $~~$ \\
            \end{tabular}
        \end{minipage}
    }
\end{table}

\endgroup

\begingroup

\setlength{\tabcolsep}{3pt}
\renewcommand{\arraystretch}{1.1}

\begin{table*}
    \centering
    \caption{Our coding of the DCMS Full guidelines (13 items) and Sub-Topics (28 items), using the SAcoding method. DCMS Full guidelines coded using guideline description (versus title, as included here). Table \ref{tab:taggingresults/DCMS} is derived from this data.}
    \label{tab:taggingoutput/DCMSSubTopic}
    \vspace{-6pt}
    \footnotesize{
        \begin{tabular}{@{}p{1.0cm}p{0.5cm}p{1.0cm}p{13.5cm}@{}}
            \toprule
   Guideline & Full & Sub-Topic  & Sub-Topic Text (from \cite{DCMScodeOfPractice}) \\
             & Code & Code       &   \\
            \toprule
            DCMS-1 & P5 &  & \protect $\bullet$~DCMS-1 title: \textbf{No default passwords}\\
            \quad 1.1 && P5& All IoT device passwords shall be unique and not resettable to any universal factory default value.  \\
            \midrule
            DCMS-2 & M1 & & \protect $\bullet$~DCMS-2 title: \textbf{Implement a vulnerability disclosure policy}\\
            \quad  2.1 && P5& All companies that provide internet-connected devices and services shall provide a public point of contact as part of a vulnerability disclosure policy in order that security researchers and others are able to report issues.  \\
            \quad  2.2 && P1& Disclosed vulnerabilities should be acted on in a timely manner.  \\
            \midrule
            DCMS-3 & M1& & \protect $\bullet$~DCMS-3 title: \textbf{Keep software updated}\\
            \quad  3.1 && O1a& Software components in internet-connected devices should be securely updateable.  \\
            \quad  3.2 && P1& Updates shall be timely and {[}...{]}  \\
            \quad  3.3 && O1a& {[}Updates{]} should not impact on the functioning of the device.  \\
            \quad  3.4 && P5& An end-of-life policy shall be published for end-point devices which explicitly states the minimum length of time for which a device will receive software updates and the reasons for the length of the support period. The need for each update should be made clear to consumers and an update should be easy to implement.  \\
            \quad  3.5 && O1a & For constrained devices that cannot physically be updated, the product should be isolatable and replaceable.  \\
            \midrule
            DCMS-4 & M1 & & \protect $\bullet$~DCMS-4 title: \textbf{Securely store credentials and security-sensitive data}\\
            \quad  4.1 && P1& Any credentials shall be stored securely within services and on devices.  \\
            \quad  4.2 && P5& Hard-coded credentials in device software are not acceptable.  \\
            \midrule
            DCMS-5 & M1 & & \protect $\bullet$~DCMS-5 title: \textbf{Communicate securely}\\
            \quad  5.1 && P1& Security-sensitive data, including any remote management and control, should be encrypted in transit, appropriate to the properties of the technology and usage.  \\
            \quad  5.2 && O1a& All keys should be managed securely.  \\
            \midrule
            DCMS-6 & N2 & & \protect $\bullet$~DCMS-6 title: \textbf{Minimise exposed attack surfaces}\\
            \quad  6.1 && N2& All devices and services should operate on the ‘principle of least privilege’; unused ports should be closed, hardware should not unnecessarily expose access, services should not be available if they are not used and code should be minimized to the functionality necessary for the service to operate. Software should run with appropriate privileges, taking account of both security and functionality  \\
            \midrule
            DCMS-7 & M1 & & \protect $\bullet$~DCMS-7 title: \textbf{Ensure software integrity}\\
            \quad  7.1 && P1& Software on IoT devices should be verified using secure boot mechanisms.  \\
            \quad  7.2 && P5& If an unauthorised change is detected, the device should alert the consumer/administrator to an issue and should not connect to wider networks than those necessary to perform the alerting function.  \\
            \midrule
            DCMS-8 & M1 & & \protect $\bullet$~DCMS-8 title: \textbf{Ensure that personal data is protected}\\
            \quad  8.1 && P5& Where devices and/or services process personal data, they shall do so in accordance with applicable data protection law, such as the General Data Protection Regulation (GDPR) and the Data Protection Act 2018.  \\
            \quad  8.2 && P5& Device manufacturers and IoT service providers shall provide consumers with clear and transparent information about how their data is being used, by whom, and for what purposes, for each device and service. This also applies to any third parties that may be involved (including advertisers).  \\
            \quad  8.3 && P1& Where personal data is processed on the basis of consumers’ consent, this shall be validly and lawfully obtained, with those consumers being given the opportunity to withdraw it at any time.  \\
            \midrule
            DCMS-9 & M1 & & \protect $\bullet$~DCMS-9 title: \textbf{Make systems resilient to outages}\\
            \quad  9.1 && O1a& Resilience should be built in to IoT devices and services where required by their usage or by other relying systems, taking into account the possibility of outages of data networks and power.  \\
            \quad  9.2 && O1a& As far as reasonably possible, IoT services should remain operating and locally functional in the case of a loss of network and {[}...{]}  \\
            \quad  9.3 && O1a& {[}Devices{]} should recover cleanly in the case of restoration of a loss of power.  \\
            \quad  9.4 && P1& Devices should be able to return to a network in a sensible state and in an orderly fashion, rather than in a massive scale reconnect.  \\
            \midrule
            DCMS-10 & P1 & & \protect $\bullet$~DCMS-10 title: \textbf{Monitor system telemetry data}\\
            \quad  10.1 && P1& If telemetry data is collected from IoT devices and services, such as usage and measurement data, it should be monitored for security anomalies.  \\
            \midrule
            DCMS-11 & M1 & & \protect $\bullet$~DCMS-11 title: \textbf{Make it easy for consumers to delete personal data}\\
            \quad  11.1 && P1& Devices and services should be configured such that personal data can easily be removed from them when there is a transfer of ownership, when the consumer wishes to delete it and/or when the consumer wishes to dispose of the device.  \\
            \quad  11.2 && P2& Consumers should be given clear instructions on how to delete their personal data.  \\
            \midrule
            DCMS-12 & M1 & & \protect $\bullet$~DCMS-12 title: \textbf{Make installation and maintenance of devices easy}\\
            \quad  12.1 && P1& Installation and maintenance of IoT devices should employ minimal steps and should follow security best practice on usability.  \\
            \quad  12.2 && P2 & Consumers should also be provided with guidance on how to securely set up their device.  \\
            \midrule
            DCMS-13 & P1 & & \protect $\bullet$~DCMS-13 title: \textbf{Validate input data}\\
            \quad  13.1 && P1& Data input via user interfaces and transferred via application programming interfaces (APIs) or between networks in services and devices shall be validated.  \\
            \bottomrule
        \end{tabular}
    }
\end{table*}

\endgroup

\end{document}